\documentclass[a4paper]{jpconf}
\usepackage{graphicx}
\begin{document}
\title{Magnetic anisotropy of YbNi$_4$P$_2$}

\author{Cornelius Krellner$^{1,2,*}$ and Christoph Geibel$^1$}
\address{$^1$ Max Planck Institute for Chemical Physics of Solids, D-01187 Dresden, Germany }
\address{$^2$ Cavendish Laboratory, University of Cambridge, Cambridge CB3 0HE, United Kingdom.}

\ead{$^*$krellner@cpfs.mpg.de}

\begin{abstract}
We report on transport and magnetic measurements between 1.8 and 400\,K on single crystalline YbNi$_4$P$_2$, which was recently reported to be a heavy fermion system with a low lying ferromagnetic transition at $T_C=0.17$\,K, based on data from polycrystals. The tetragonal crystal structure of YbNi$_4$P$_2$ presents quasi-one-dimensional Yb chains along the $c$ direction. Here we show that at high temperatures, the magnetic anisotropy of YbNi$_4$P$_2$ is dominated by the crystal electrical field effect with an Ising-type behaviour, which gets more pronounced towards lower temperatures. The electrical resistivity also reflects the strong anisotropy of the crystal structure and favours transport along $c$, the direction of the Yb chains.  
\end{abstract}

\section{Introduction}
Within the field of strongly correlated electron systems, the discovery of remarkable phases and transitions is often tightly coupled to the growth and characterization of novel materials. Studying these emergent phenomena necessitate the preparation of high-quality single crystals to investigate the physical properties as a function of crystal orientation and momentum. YbNi$_4$P$_2$ is such a new stoichiometric compound where we recently observed a very low lying ferromagnetic (FM) transition at $T_C=0.17$\,K \cite{Krellner:2011}. Above this transition distinct deviations from the predictions of Landau-Fermi-liquid theory were observed, indicating the presence of a nearby FM quantum critical point. Among $4f$-based systems this is the clearest experimental example of FM quantum criticality \cite{Stewart:2001}, although further investigations are necessary to understand this material in more detail.

\begin{figure}[t]
\begin{center}
\includegraphics[width=0.8\textwidth]{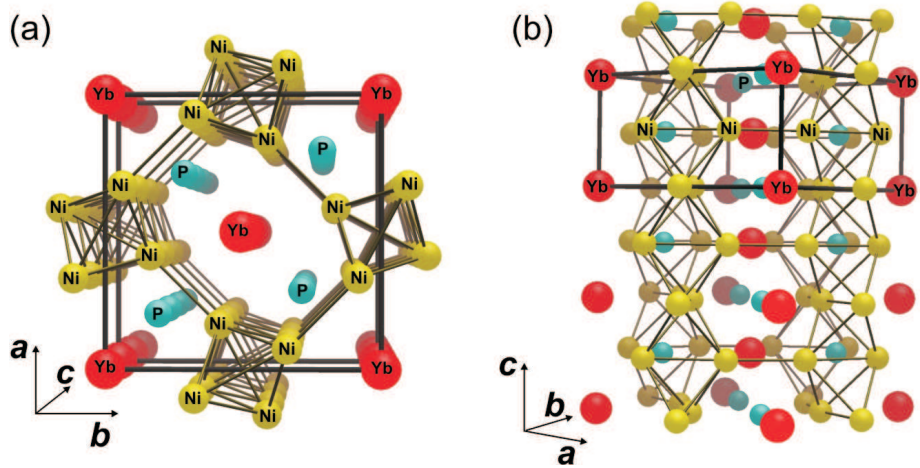}
\caption{\label{fig1}Tetragonal crystal structure of YbNi$_4$P$_2$. (a) Stereoscopic view along $c$ with the Yb chains
located in the channels between chains of edge-connected Ni tetrahedra. (b) Rotated view, with the $c$ direction along the vertical axis, the Yb chains are well separated from each other in the $ab$ plane.}
\end{center}
\end{figure}

A further intriguing property of YbNi$_4$P$_2$ is the quasi-one-dimensionality, with  Yb and Ni-tetrahedra chains along the $c$ direction of the tetragonal crystal structure  \cite{Chikhrij:1986}. These chains are well separated in the $ab$ plane, which is due to the small $c/a\approx 0.5$ ratio.  In Fig.~\ref{fig1}a the structure is presented using a stereoscopic view along $c$ with the Yb chains located in the channels between chains of edge-connected Ni tetrahedra. Turning the structure by 90$^{\circ}$ one clearly sees the Yb chains running from bottom to the top in Fig.~\ref{fig1}b. The quasi-one-dimensional anisotropy was theoretically confirmed by uncorrelated band-structure calculations, which reveal that the three main Fermi surfaces all have a predominantly one-dimensional character \cite{Krellner:2011}.

So far, the anisotropy of the physical properties of YbNi$_4$P$_2$ have not been investigated, as the previous studies were performed on polycrystalline samples \cite{Krellner:2011, Deputier:1997}. In this contribution, we report on magnetic and transport measurements on single crystals between 1.8 and 400\,K, which reveal a pronounced single-ion anisotropy originating from crystal electric field (CEF) effects.

\section{Experimental}
The single crystals of YbNi$_4$P$_2$ were grown in a Ni-rich self-flux at 1400$^{\circ}$C in a closed Ta container. The excess flux was removed by centrifuging at high temperatures followed by etching in diluted nitric acid. The crystals grow without preferred growth direction and were oriented and subsequently cut using back-scattering Laue and a conventional diamond saw, respectively. Several powder X-ray diffraction patterns of crushed single crystals confirmed the formation of YbNi$_4$P$_2$ with the ZrFe$_4$Si$_2$ structure type (space group: $P4_2/mnm$; $Z=2$). The lattice and structure parameters were found to be in good agreement with the reported structure data \cite{Kuzma:2000} and are given in detail in Ref.~\cite{Krellner:2011}. In addition, energy dispersive X-ray spectra performed on a scanning electron microscope (Philips XL30) proved the formation of single phase YbNi$_4$P$_2$. However, the exact stoichiometry could not be refined with the latter method, as the L-line of Yb and the K-line of Ni nearly completely overlap in the energy resolved spectra. The dc-susceptibility and  magnetization data were recorded using a Quantum Design (QD) VSM SQUID equipped with a 7 T Magnet. Resistivity measurements were done using standard ac four-probe geometry in a QD PPMS.

\section{Results and Discussion}
\begin{figure}[t]
\includegraphics[width=0.62\textwidth]{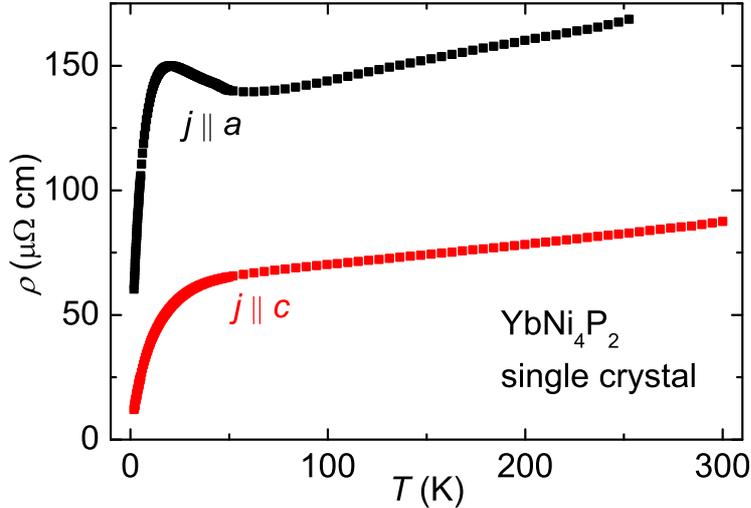}
\begin{minipage}[b]{14pc}\caption{\label{fig2}Temperature dependence of the electrical resistivity with the current along the $a$ (black) and the $c$ (red) direction. The resistivity along the Yb chains ($j\| c$) is distinctevely smaller than along $a$ in the whole invetigated temperature range. Both directions display a pronounced drop towards lower temperatures due to the onset of coherent Kondo scattering. A pronounced Kondo maximum at $T=20$\,K is visible only for $j\| a$.}
\end{minipage}
\end{figure}

The electrical resistivity $\rho(T)$ shown in Fig.~\ref{fig2} evidences clearly the anisotropy expected from the crystal and electronic structure. Above $100$\,K, the resistivity along $c$, the direction of the Yb chains, is a factor of 2 smaller than perpendicular to the chains ($j\| a$). In this temperature range the electrical transport is dominated by phonons, reflected in a linear-in-$T$ dependence for both directions. At $1.8$\,K, the anisotropy $\rho_a/\rho_c$ amounts already to a factor of 5. Furthermore, the temperature dependences below $50$\,K are strongly dependent on the direction of current flow. For the easy transport direction ($j\| c$), a pronounced drop is observed below $T=30$\,K, characteristic of coherent Kondo scattering. For $j\| a$, $\rho(T)$ increases  below 50\,K, presents a well defined maximum at $T=20$\,K and then also strongly decreases towards lower temperatures. Such distinct anisotropic behaviour in this $T$ range is typical for Kondo-lattice systems and results from the single-ion anisotropy induced by the CEF \cite{Kashiba:1986}.

\begin{figure}[t]
\begin{center}
\includegraphics[width=\textwidth]{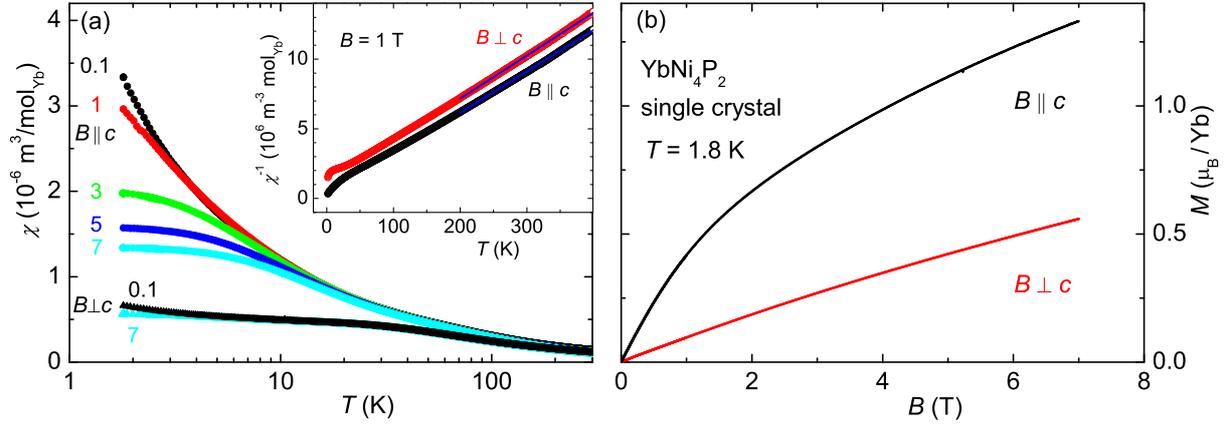}
\caption{\label{fig3}Magnetic anisotropy of YbNi$_4$P$_2$. (a) $\chi(T)$ for fields parallel (circles) and perpendicular (triangles) to the $c$ direction. In the inset, $\chi^{-1}(T)$ is shown for both directions at $B=1$\,T. (b) Isothermal magnetization as a function of applied magnetic field at $T=1.8$\,K.} 
\end{center}
\end{figure}

The anisotropy of the susceptibility, $\chi(T)$, above 1.8\,K is similarly dominated by the single-ion effects, although less pronounced at high temperatures. The inverse susceptibility presents a perfect Curie-Weiss behaviour between 200 and 400\,K (blue lines),  shown in the inset of Fig.~\ref{fig3}a. The effective moments for magnetic field along $c$, $\mu_{eff}^c=4.63\,\mu_B$, and perpendicular to $c$, $\mu_{eff}^{ab}=4.54\,\mu_B$, match that of the free Yb$^{3+}$ moment ($4.54\,\mu_B$). The anisotropy at high temperatures is reflected in different Weiss temperatures, $\Theta_W^c=-7$\,K and $\Theta_W^{ab}=-33$\,K, for the applied field parallel to $c$ and perpendicular to $c$, respectively. Below $50$\,K, the anisotropy of $\chi(T)$ gets more pronounced with a much stronger increase for $B\| c$ towards lower $T$ and a tendency to saturation for $B\perp c$. The absolute values at $T=1.8\,$K and $B=0.1$\,T are a factor of 5 larger for $B\| c$ than for $B\perp c$, i.e., $\chi^c/\chi^{ab}(0.1$\,T$)\approx 5$. From the polycrystalline data it is known, that $\chi(T)$ further increases towards lower temperatures and nearly diverges at $T_C$ \cite{Krellner:2011}. Presently, we cannot state anything about the anisotropy of the inter-ionic magnetic exchange interaction, which most likely will dominate the susceptibility data in small fields towards $T_C$. Therefore, the anisotropy of $\chi(T)$ needs to be studied on these single crystals down to the mK range in the future.

The magnetization, $M(B)$, as function of applied magnetic field at 1.8\,K is presented for both directions in Fig.~\ref{fig3}b. For $B\| c$, we observe a strongly curved $M(B)$, in agreement with the strong field dependence of the susceptibility measured on the same sample. At 7\,T, $M(B)$ still increases towards higher fields and the polarized moment amounts to $1.33\,\mu_B$. For $B\perp c$, the $M(B)$ curve is nearly linear with a much smaller  polarized moment of $0.56\,\mu_B$ at 7\,T. 

These observations of the magnetic properties above 1.8\,K can be well understood in terms of the CEF, which leads to a pronounced single-ion anisotropy with Ising-type behaviour, i.e., the magnetic easy direction is the $c$ axis of the tetragonal unit cell. On the basis of the present results, a complete determination of the CEF parameters cannot be given, which would need further experimental input. However, some constraints of the CEF scheme can be already drawn from the existing data. Specific heat measurements on the polycrystal down to 40\,mK revealed the entropy in zero magnetic field \cite{Krellner:2011}, from which we can conclude that the CEF ground state of YbNi$_4$P$_2$ is a doublet, as the entropy at 4\,K amounts to $0.5R\ln 2$. Furthermore, we can estimate by means of the minimum in the thermopower at 35\,K, the maximum at 20\,K in $\rho(T)$ along the $a$ direction and the pronounced deviation from the Curie-Weiss behaviour in the susceptibility below 50\,K, that the first excited CEF doublet lies roughly between 10 to 30\,K above the CEF ground state. Finally, the fact that we recover the full effective moment in the Curie-Weiss law between 200 and 400\,K gives some evidence that the complete splitting of the 4 Kramers doublets of the Yb$^{3+}$ is not much larger than 400\,K. However, inelastic neutron measurements are underway to determine the CEF scheme more accurately.
 
\section{Conclusion}
In summary, we succeeded in growing single crystals of YbNi$_4$P$_2$ large enough to determine the anisotropy by means of magnetic susceptibility and resistivity measurements. We showed that the anisotropy above 1.8\,K is dominated by  crystalline electric field effects with an Ising-type single-ion ground state. This is reflected in a much higher polarized moment at 1.8\,K for magnetic field along the $c$ direction compared to the in-plane moment. The resistivity presents large anisotropy in the whole investigated temperature range and reveals more efficient transport along the crystallographic $c$ direction, the direction of the quasi-one-dimensional Yb chains. 
Further investigations to lower temperatures on these single crystals are needed to reveal the anisotropy of the magnetic exchange and the direction of the moment within the ferromagnetic ordered state.

\section*{Acknowledgments}
The authors thank U. Burkhardt and P. Scheppan for chemical analysis of the samples, as well as C. Bergmann, N. Caroca-Canales and R. Weise for technical assistance in sample preparation. 
This work is partially supported through DFG Research Unit 960 and the SPP 1458.

\end{document}